\documentclass[11pt]{amsart}

\newtheorem{Theorem}{Theorem}

\newtheorem{Lemma}{Lemma}

\newcommand{\N}{\ensuremath{{\mathrm{I\!N}}}}
\newcommand{\Z}{\ensuremath{{\mathsf{Z\!\!Z}}}}
\newcommand{\C}{\ensuremath{{\mathrm{C\hspace{-1.7mm}\rule{0.3mm}{2.6mm}\;}}}}
\newcommand{\R}{\ensuremath{{\mathrm{I\!R}}}}
\newcommand{\Id}{{\mathrm{I_d}}}

\newcommand{\Rem}{{\mathrm{Re}~}}
\newcommand{\Imm}{{\mathrm{Im}~}}
\newcommand{\trip}[1]{{|\kern -1pt\|#1\|\kern -1pt|}}

\begin{document}

\title[Pure Point and Energy Instability]{A Floquet Operator with Pure Point
Spectrum and Energy
Instability}
\author{C\'{e}sar R. de Oliveira}
\thanks{CRdeO was partially supported by CNPq (Brazil).}
\address{Departamento de Matem\'{a}tica -- UFSCar, S\~{a}o Carlos, SP, 13560-970
Brazil}
\email{oliveira@dm.ufscar.br}
\author{Mariza S. Simsen}
\thanks{MSS was supported by CAPES (Brazil).}
\address{Departamento de Matem\'{a}tica -- UFSCar, S\~{a}o Carlos, SP, 13560-970
Brazil}
\email{mariza@dm.ufscar.br}
 \subjclass{81Q10 (11B,47B99)}

\begin{abstract} An example of Floquet operator with purely point spectrum and
energy instability is
presented. In the unperturbed energy eigenbasis its eigenfunctions are exponentially
localized.
\end{abstract}
\maketitle

\section{Introduction} It is not immediate whether a self-adjoint operator $H$ with
purely point spectrum
 implies absence of transport under the time evolution $U(t)=e^{-iHt}$; in fact, it
is currently known
examples of Schr\"odinger operators with such kind of  spectrum and transport. In
case of tight-binding
models on $l^2(\N)$ the transport is usually probed by the moments of order $m>0$ of
the position
operator $X e_k=k e_k$, that is,
\begin{equation}\label{MomentEquation} X^m=\sum_{k\in\N}k^m\langle e_k,\cdot\rangle
e_k,
\end{equation} where $e_k(j)=\delta_{kj}$ (Kronecker delta) is the canonical basis
of $l^2(\N)$. Analogous
definition applies for $l^2(\Z)$ and even higher dimensional spaces. Then, by
definition, {\em transport} at
$\psi$, also called {\em dynamical instability} or {\em dynamical delocalization},
occurs if for some $m$
the function
\begin{equation}\label{funcX} t\mapsto \langle \psi(t),X^m\psi(t)\rangle,\quad
\psi(t):=U(t)\psi,
\end{equation} is unbounded. If for all $m>0$ the corresponding functions are
bounded, one has {\em
dynamical stability}, also called {\em dynamical localization}. 

The first rigorous example of a
Schr\"odinger operator with purely point spectrum and dynamical instability has
appeared
in~\cite{dRJLS}, Appendix~2, what the authors have called ``A Pathological
Example;'' in this case the
tight binding Schr\"odinger operator
$h$  on
$l^2(\N)$ with a Dirichlet condition at $n=-1$ was
\[(hu)(n)=u(n+1)+u(n-1)+v(n)u(n)\] with potential 
\begin{equation}\label{exemDeles} v(n)=3\cos(\pi\alpha n+\theta)+\lambda\delta_{n0},
\end{equation} that is, rank one perturbations of an instance of the almost Mathieu
operator. An irrational
number
$\alpha$ was constructed  so that for a.e.\ $\theta\in[0,2\pi)$ and a.e.\
$\lambda\in[0,1]$ the corresponding self-adjoint operator $h$ has purely point 
spectrum with
dynamical instability at $e_0$ (throughout, the term ``a.e.''\ without specification
means with respect to
the Lebesgue measure under consideration). More precisely, it was shown that for all
$\epsilon>0$
\[
\limsup_{t\to\infty} \frac{1}{t^{2-\epsilon}} \langle
\psi(t),X^2\psi(t)\rangle = \infty,\quad
\psi(0)=e_0.
\] Compare with the absence of ballistic motion for point spectrum Hamiltonians
\cite{S0}
\[
\lim_{t\to\infty} \frac{1}{t^{2}} \langle
\psi(t),X^2\psi(t)\rangle = 0.
\]  

Additional
examples of this behavior are known, even for random potentials, but with a strong
local correlations
\cite{JSS}, as for the random dimer model in the Schr\"odinger case; there is also
an adaptation
\cite{dOP1} for the random Bernoulli Dirac operator with no correlation in the
potential, although for the
massless case.

The time evolution of a quantum system with time-dependent Ha\-mil\-to\-nian is
given by a  strongly
continuous family of unitary operators $U(t,r)$ (the propagator). For an initial
condition
$\psi_0$ at $t=0$, its time evolution is given by $U(t,0)\psi_0$. If the Hamiltonian
is time-periodic with
period
$T$, then 
\[
U(t+T,r+T)=U(t,r),\quad\forall t,r,
\] and we have the Floquet operator $U_F:= U(T,0)$ defined as the evolution
generated by the Hamiltonian over a period.

Quantum systems governed by a time periodic Hamiltonian have their dynamical
stability often characterized
in terms of the spectral properties of the corresponding Floquet operator. As in the
autonomous case, the
presence of continuous spectrum is a signature of unstable quantum systems; this is
a consequence of the
famous RAGE theorem, firstly proved for the autonomous case~\cite{RS} and then for
time-periodic
Hamiltonians~\cite{EV}. In principle, a Floquet operator with purely point spectrum
would imply
``stability,'' but one should be alerted by the above mentioned ``pathological''
examples in the autonomous
case.

  In this work we give an example of a Floquet operator with purely point spectrum
and ``energy
instability,'' which can be considered the partner concept of dynamical instability
in case
of autonomous systems. We shall consider a particular choice in the family of
Floquet operators  studied
in~\cite{BHJ}; such operators describe the quantum dynamics of certain interesting
physical models
(see~\cite{BB,BHJ} and references therein), and display a band structure with
respect to an orthogonal
basis
$\{\varphi_k\}$ of $l^2(\N)$ or $l^2(\Z)$, constructed as eigenfunctions of an
unperturbed energy operator.
There are some conceptual differences with respect to the autonomous case mentioned
before, since now the
momentum
$X^m$ is defined in the energy space
\begin{equation}\label{MomentEquationEnergy} X^m=\sum_{k\ge
1}k^m\langle\varphi_k,\cdot\rangle\varphi_k,
\end{equation} instead of the ``physical space'' $\N$. Thus, if for all $m>0$ the
functions
\begin{equation}\label{funcXFloquet} n\mapsto \langle
\psi(n),X^m\psi(n)\rangle,\quad \psi(n):=U_F^n\psi,
\quad n\ge0,
\end{equation}
are
bounded we say there is {\em energy stability} or {\em energy localization}, while
if at least one of
them is unbounded we say the system presents {\em energy instability} or  {\em
energy delocalization}; the
latter reflects a kind of ``resonance.''

Our construction is a fusion of the Floquet operator studied in~\cite{BHJ}, now
under suitable additional
rank one perturbations, and the arguments presented in~\cite{dRJLS} for model
(\ref{exemDeles}). For
suitable values of parameters we shall get the following properties:
\begin{enumerate}
\item The resulting unitary operator $U_{\lambda}(\beta,\theta)^{+}$ (after the rank
one perturbation; see
Eq.~(\ref{FR1P}))  still belongs to the family of Floquet operators considered
in~\cite{BHJ}.
\item $U_{\lambda}(\beta,\theta)^{+}$ has purely point spectrum with exponentially
localized eigenfunctions.
\item The time evolution along the Floquet operator $U_{\lambda}(\beta,\theta)^{+}$
of the initial condition
$\varphi_1$ presents energy instability.
\end{enumerate}

$U_{\lambda}(\beta,\theta)^{+}$ will be obtained as a rank one perturbation of the
almost periodic class of
operators studied in the Section 7 of~\cite{BHJ} (we describe them ahead). In order
to prove purely point
spectrum, we borrow an argument from~\cite{HJS} that was used to prove localization
for random unitary
operators, and it combines spectral averaging and positivity of the Lyapunov
exponent with polynomial
boundedness of generalized eigenfunctions. In order to get dynamical instability,
although we adapt ideas
of~\cite{dRJLS},  we underline that some results needed completely different proofs
and they are not
entirely trivial. 

It is worth mentioning that in \cite{SimonCVM} a form of dynamical stability was
obtained for discrete
evolution along some Floquet operators (CMV matrices) related to random Verblunsky
coefficients.

This paper is organized as follows. In Section~\ref{FloquetSection} we present the
model of Floquet
operator  we shall consider, some preliminary results  and the main result is stated
in Theorem~\ref{MainTheorem}. In Section~\ref{SpectrumSection} we shall prove that
our Floquet operator is
pure point. Section~\ref{InstabilitySection} is devoted to the proof of dynamical
instability.

\

\section{The Floquet Operator}
\label{FloquetSection} We briefly recall the construction of the Floquet operator
introduced in~\cite{BHJ}
based on the physical model discussed in~\cite{BB}. The separable Hilbert space is
$l^2(\Z)$ and
$\{\varphi_k\}_{k\in\Z}$ denote its canonical basis. Consider the set of
$2\times2$ matrices defined for any $k\in\Z$ by
\[S_k=e^{-i\theta_k}\left(\begin{array}{cc} re^{-i\alpha_k} & ite^{i\gamma_k}\\
ite^{-i\gamma_k} &
re^{i\alpha_k}
\end{array}\right)\] parameterized by the phases
$\alpha_k,\;\gamma_k,\;\theta_k$ in the torus $\mathbb{T}$ and the real parameters
$t,r$, the reflection
and transition coefficients, respectively, linked by $r^2+t^2=1$. Then, let $P_j$ be
the orthogonal
projection onto the span of
$\varphi_j,\;\varphi_{j+1}$ in $l^2(\Z)$, and let $U_e,\;U_o$ be two $2\times2$
block diagonal unitary
operators on $l^2(\Z)$ defined by
\[U_e=\sum_{k\in\Z}P_{2k}S_{2k}P_{2k}\quad\textrm{and}\quad
U_o=\sum_{k\in\Z}P_{2k+1}S_{2k+1}P_{2k+1}.\] The
matrix representation of $U_e$ in the canonical basis is
\[U_e=\left(\begin{array}{ccccc} \ddots & & & & \\ & S_{-2} & & & \\
 & & S_0 & & \\ & & & S_2 & \\ & & & & \ddots \end{array}\right),\] and similarly
for $U_o$, with $S_{2k+1}$
in place of $S_{2k}$. The Floquet operator $U$ is defined by
\[U=U_oU_e,\] such that, for any $k\in\Z$,
\begin{eqnarray}\label{OperatorEquation} U\varphi_{2k} &=&
irte^{-i(\theta_{2k}+\theta_{2k-1})}
e^{-i(\alpha_{2k}-\gamma_{2k-1})}\varphi_{2k-1}\nonumber\\  & &
+r^2e^{-i(\theta_{2k}+\theta_{2k-1})}
e^{-i(\alpha_{2k}-\alpha_{2k-1})}\varphi_{2k}\nonumber\\ & &
+irte^{-i(\theta_{2k}+\theta_{2k+1})}
e^{-i(\gamma_{2k}+\alpha_{2k+1})}\varphi_{2k+1}\nonumber\\ & &
-t^2e^{-i(\theta_{2k}+\theta_{2k+1})}
e^{-i(\gamma_{2k}+\gamma_{2k+1})}\varphi_{2k+2}\nonumber\\ U\varphi_{2k+1} &=&
-t^2e^{-i(\theta_{2k}+\theta_{2k-1})}
e^{i(\gamma_{2k}+\gamma_{2k-1})}\varphi_{2k-1}\\ & &
+irte^{-i(\theta_{2k}+\theta_{2k-1})}
e^{i(\gamma_{2k}+\alpha_{2k-1})}\varphi_{2k}\nonumber\\ & &
+r^2e^{-i(\theta_{2k}+\theta_{2k+1})}
e^{i(\alpha_{2k}-\alpha_{2k+1})}\varphi_{2k+1}\nonumber\\ & &
+irte^{-i(\theta_{2k}+\theta_{2k+1})}
e^{i(\alpha_{2k}-\gamma_{2k+1})}\varphi_{2k+2}\nonumber
\end{eqnarray}

The extreme cases where $rt=0$ are spectrally trivial; in case
$t=0$, $r=1$, $U$ is pure point and if $t=1$, $r=0$, $U$ is purely absolutely
continuous (Proposition 3.1
in \cite{BHJ}). From now on we suppose $0<r,t<1$.

For the eigenvalue equation
\[U\psi=e^{iE}\psi\]
\[\psi=\sum_{k\in\Z}c_k\varphi_k,\qquad c_k,E\in\C,
\] one gets the following relation between  coefficients
\[\left(\begin{array}{c} c_{2k} \\ c_{2k+1}
\end{array}\right)=T_k(E) \left(\begin{array}{c} c_{2k-2} \\ c_{2k-1}
\end{array}\right),
\] where the matrix $T_k(E)$ has elements
\begin{eqnarray}\label{TransferEquation}
T_k(E)_{11}&=&-e^{-i\left(E+\gamma_{2k-1}+\gamma_{2k-2}+\theta_{2k-1}
+\theta_{2k-2}\right)},\nonumber\\
T_k(E)_{12}&=&i\frac{r}{t}\left(e^{-i\left(E+\gamma_{2k-1}-\alpha_{2k-2}
+\theta_{2k-1}+\theta_{2k-2}\right)}-e^{-i\left(\gamma_{2k-1}-\alpha_{2k-1}\right)}\right),
\nonumber\\ T_k(E)_{21} &=&
i\frac{r}{t}\Big(e^{-i\left(\theta_{2k-2}-\theta_{2k}+\gamma_{2k}
+\gamma_{2k-1}+\gamma_{2k-2}+\alpha_{2k-1}\right)}\nonumber\\ & & -e^{-i\left(E+
\theta_{2k-2}+\theta_{2k-1}+\gamma_{2k}+\gamma_{2k-1}+\gamma_{2k-2}
+\alpha_{2k}\right)}\Big),\\
T_k(E)_{22} &=& -\frac{1}{t^2}e^{i\left(E+\theta_{2k}+\theta_{2k-1}-\gamma_{2k}
-\gamma_{2k-1}\right)}\nonumber\\ & &
+\frac{r^2}{t^2}e^{-i\left(\gamma_{2k}+\gamma_{2k-1}\right)}\left(e^{i\left(\theta_{2k}-
\theta_{2k-2}
+\alpha_{2k-2}-\alpha_{2k-1}\right)}+e^{-i\left(\alpha_{2k}-\alpha_{2k-1}\right)}\right)
\nonumber\\ & & -\frac{r^2}{t^2}e^{-i\left(E+\theta_{2k-2}+\theta_{2k-1}+\gamma_{2k}
+\gamma_{2k-1}+\alpha_{2k}-\alpha_{2k-2}\right)}\nonumber
\end{eqnarray} and \[\det
T_k(E)=e^{-i(\theta_{2k-2}-\theta_{2k}+\gamma_{2k}+2\gamma_{2k-1}+\gamma_{2k-2})}.\]

Given coefficients $(c_0,c_1)$,  for any $k\in\N^*$ one has
\[\left(\begin{array}{c} c_{2k} \\ c_{2k+1}
\end{array}\right)=T_k(E)\ldots T_2(E)T_1(E)\left(\begin{array}{c} c_0 \\ c_1
\end{array}\right),\]
\[\left(\begin{array}{c} c_{-2k} \\ c_{-2k+1}
\end{array}\right)=T_{-k+1}(E)^{-1}\ldots T_{-1}(E)^{-1}T_0(E)^{-1}
\left(\begin{array}{c} c_0 \\ c_1
\end{array}\right).\]

In the physical setting \cite{BB}, the natural Hilbert space is
$l^2(\N^*)$, with $\N^*$ the set of positive integers, and the definition according
with~\cite{BHJ} of the
Floquet operator, denoted by
$U^{+}$, is
\begin{eqnarray}\label{PositiveEquation}
U^{+}\varphi_1&=&re^{-i\left(\theta_0+\theta_1\right)}e^{-i\alpha_1}\varphi_1+
ite^{-i\left(\theta_0+\theta_1\right)}e^{-i\gamma_1}\varphi_2,\nonumber\\
U^{+}\varphi_k&=&U\varphi_k,\;k>1
\end{eqnarray} with $U\varphi_k$ as in (\ref{OperatorEquation}). In this case the
eigenvalue equation is
\[ U^{+}\psi=e^{iE}\psi
\] with
$\psi=\sum_{k=1}^{\infty}c_k\varphi_k$. Then starting from
$c_2,c_3$, we have
\[\left(\begin{array}{c} c_{2k} \\ c_{2k+1}\end{array}\right)=T_k(E)\ldots T_2(E)
\left(\begin{array}{c} c_2 \\ c_3\end{array}\right), \qquad k=2,3,\ldots\] where the
transfer matrices
$T_k(E)$ are given by (\ref{TransferEquation}), along with the additional one
\[\left(\begin{array}{c} c_2 \\ c_3\end{array}\right)=c_1
\left(\begin{array}{c} a_1(E) \\ a_2(E)\end{array}\right),\] where
\begin{eqnarray*}
a_1(E)&=&\frac{i}{t}\left(e^{-i\left(E+\gamma_1+\theta_1+\theta_0\right)}-r
e^{-i\left(\gamma_1-\alpha_1\right)}\right)\\ a_2(E) &=&
-\frac{1}{t^2}e^{i\left(E+\theta_2+\theta_1-\gamma_2 -\gamma_1\right)}\\ & &
+\frac{r}{t^2}e^{-i\left(\gamma_2+\gamma_1\right)}\left(e^{i\left(\theta_2-
\theta_0-\alpha_1\right)}+re^{-i\left(\alpha_2-\alpha_1\right)}\right)\\ & &
-\frac{r}{t^2}e^{-i\left(E+\theta_0+\theta_1+\gamma_2+\gamma_1+\alpha_2\right)}
\end{eqnarray*}

For further details and generalizations of this class of unitary operators, we refer
the reader
to~\cite{BHJ,J,J1,HJS}. In particular, when the phases are i.i.d.\ random variables,
it was proved to hold
in the unitary case typical results obtained for discrete one-dimensional random
Schr\"{o}dinger operators.
For example, the availability of a transfer matrix formalism to express generalized
eigenvectors allows to
introduce a Lyapunov exponent, to prove a unitary version of Oseledec's Theorem and
of Ishii-Pastur Theorem
(and get absence of absolutely continuous spectrum in some cases).

Our main interest is on the almost periodic example 
\[ U\equiv U(\{\theta_k\},\{\alpha_k\},\{\gamma_k\}),
\] where the phases
$\alpha_k$ are taken as constants, $\alpha_k=\alpha\;\;\forall k\in\Z$, while the
$\gamma_k$'s are
arbitrary and can be replaced by $(-1)^{k+1}\alpha$ (see Lemma 3.2 in~\cite{BHJ}).
The almost periodicity
due to the phases $\theta_k$ defined by
$\theta_k=2\pi\beta k+\theta,$ where $\beta\in\R$, and
$\theta\in[0,2\pi)$. We denote $U$ above by $U=U(\beta,\theta)$ and then for any
$k\in\Z$ (see
(\ref{OperatorEquation}))
\begin{eqnarray}\label{ThetaEquation} U(\beta,\theta)\varphi_{2k} &=&
irte^{-i(2\pi\beta(4k-1)+2\theta)}\varphi_{2k-1}\nonumber\\ & &
+r^2e^{-i(2\pi\beta(4k-1)+2\theta)}\varphi_{2k}\nonumber\\ & &
+irte^{-i(2\pi\beta(4k+1)+2\theta)}\varphi_{2k+1}\nonumber\\ & &
-t^2e^{-i(2\pi\beta(4k+1)+2\theta)}\varphi_{2k+2}\nonumber\\
U(\beta,\theta)\varphi_{2k+1} &=&
-t^2e^{-i(2\pi\beta(4k-1)+2\theta)}\varphi_{2k-1}\\ & &
+itre^{-i(2\pi\beta(4k-1)+2\theta)}\varphi_{2k}\nonumber\\ & &
+r^2e^{-i(2\pi\beta(4k+1)+2\theta)}\varphi_{2k+1}\nonumber\\ & &
+itre^{-i(2\pi\beta(4k+1)+2\theta)}\varphi_{2k+2}\nonumber
\end{eqnarray} Let $U(\beta,\theta)^{+}$ be the corresponding operator on
$l^2(\N^*)$ defined by (\ref{PositiveEquation}). The following result was proved
in~\cite{BHJ}.

\

\begin{Theorem} \label{SpectrumTheorem} (i) For $\beta$ rational and each
$\theta\in[0,2\pi)$,
$U(\beta,\theta)$ is purely absolutely continuous,
$\sigma_{\mathrm{sc}}(U(\beta,\theta)^{+})=\emptyset$,
$\sigma_{\mathrm{ac}}(U(\beta,\theta)^{+})=\sigma_{\mathrm{ac}}(U(\beta,\theta))$
and the point spectrum of
$U(\beta,\theta)^{+}$ consists of finitely many simple eigenvalues in the resolvent
set of
$U(\beta,\theta)$.
\newline (ii) Let $T_k^{\theta}(E)$ be the transfer matrices at
$E\in\mathbb{T}$ corresponding to
$U(\beta,\theta)$. For $\beta$ irrational, the Lyapunov exponent
$\gamma(E)$ satisfies, for almost all $\theta$,
\[\gamma_{\theta}(E)=\lim_{N\rightarrow\infty}\frac{\ln
\|\prod_{k=1}^NT_k^{\theta}(E)\|}{N}\geq\ln\frac{1}{t^2}>0,\] and so
$\sigma_{ac}(U(\beta,\theta))=\emptyset$. The same is true for
$U(\beta,\theta)^{+}$.
\end{Theorem}

\

Finally, we introduce our study model. We consider a rank one perturbation of
$U(\beta,\theta)^{+}$,
$\lambda\in[0,2\pi)$ (see also \cite{C})
\begin{equation}\label{FR1P}
U_{\lambda}(\beta,\theta)^{+}:= U(\beta,\theta)^{+}e^{i\lambda P_{\varphi_1}}
=U(\beta,\theta)^{+}\left(\Id+(e^{i\lambda}-1)P_{\varphi_1}\right),
\end{equation} where
$P_{\varphi_1}(\cdot)=\langle\varphi_1,\cdot\rangle\varphi_1$. We observe that
\[U(\beta,\theta)^{+}\equiv
U^{+}\left(\{\theta_k\}_{k=0}^{\infty},\{\alpha_k\}_{k=1}^{\infty},
\{\gamma_k\}_{k=1}^{\infty}\right)\] and
$U_{\lambda}(\beta,\theta)^{+}\equiv
U^{+}\left(\{\tilde{\theta_k}\}_{k=0}^{\infty},\{\tilde{\alpha_k}\}_{k=1}^{\infty},
\{\tilde{\gamma_k}\}_{k=1}^{\infty}\right)$ where
$\tilde{\theta_0}=\theta_0-\lambda$ and
$\tilde{\theta_k}=\theta_k$, $\tilde{\alpha_k}=\alpha_k$,
$\tilde{\gamma_k}=\gamma_k$ for $k\ge1$. Hence, the perturbed operator
$U_{\lambda}(\beta,\theta)^{+}$  also belongs to the family of Floquet operators
studied in~\cite{BHJ}.
Note also that the Lyapunov exponent is independent on the parameter $\lambda$.

We can now state our main result:

\

\begin{Theorem} \label{MainTheorem} (i) For $\beta$ irrational,
$U_{\lambda}(\beta,\theta)^{+}$ has only point spectrum for a.e.\
$\theta,\;\lambda\in[0,2\pi)$, and in the basis $\{\varphi_k\}$ its eigenfunctions
decay exponentially.
\newline (ii)
$\beta$ can be chosen irrational so that
\[\limsup_{n\rightarrow\infty}\frac{\|X\left(U_{\lambda}(\beta,\theta)^{+}\right)^n
\varphi_1\|^2} {F(n)}=\infty,\] for all $\theta\in[0,2\pi)$ and any
$\lambda\in[\frac{\pi}{6},\frac{\pi}{2}]$, where
$F(n)=\frac{n^2}{\ln(2+n)}$ and $X$ is the moment of order $m=1$ given by
(\ref{MomentEquationEnergy}).
\end{Theorem}

\

\noindent \textit{Remarks.} 1. Joining up (i) and (ii) of the theorem above we
proved that for some $\beta$
irrational, for a.e.\ $\theta\in[0,2\pi)$ and
$\lambda\in[\frac{\pi}{6},\frac{\pi}{2}]$,
$U_{\lambda}(\beta,\theta)^{+}$ has pure point spectrum and the function
\[ n\mapsto \left\langle \left(U_{\lambda}(\beta,\theta)^{+}\right)^n
\varphi_1,X^2\left(U_{\lambda}(\beta,\theta)^{+}\right)^n
\varphi_1\right\rangle\] is unbounded. That is, we have pure point spectrum and
dynamical instability.
\newline 2. One can modify the proof to replace the logarithm function
$f(n)=\ln(2+n)$ for any monotone sequence
$f$ with
$\lim_{n\rightarrow\infty}f(n)=\infty$.

\

\section{Pure Point Spectrum}
\label{SpectrumSection} In this section we prove  part (i) of
Theorem~\ref{MainTheorem}. We need a
preliminary lemma.

\

\begin{Lemma}\label{CyclicLemma} For any $\beta$ and $\theta$, the vector
$\varphi_1$ is cyclic for
$U(\beta,\theta)^{+}$.
\end{Lemma}
\begin{proof} Fix $\beta$ and $\theta$. We indicate that any vector
$\varphi_k$,
$k\in\N^*$ can be written as a linear combination of the vectors
$\left(U(\beta,\theta)^{+}\right)^n\varphi_1$, $n\in\Z$. Since
$U(\beta,\theta)^{+}\varphi_1=re^{-i\left(2\pi\beta+2\theta\right)}
e^{-i\alpha}\varphi_1+
ite^{-i\left(2\pi\beta+2\theta\right)}e^{-i\alpha}\varphi_2$ then
\begin{equation}\label{Eq1}
\varphi_2=-\frac{i}{t}e^{i\left(2\pi\beta+2\theta\right)}e^{i\alpha}
U(\beta,\theta)^{+}\varphi_1+\frac{ir}{t}\varphi_1.
\end{equation} Now
\begin{equation}\label{Eq2}
\left(U(\beta,\theta)^{+}\right)^{-1}\varphi_1=a_1\varphi_1+a_2\varphi_2+a_3\varphi_3,
\end{equation} where $a_1,\;a_2$ and $a_3$ are nonzero complex numbers. Thus, using
(\ref{Eq1}) and
(\ref{Eq2}), suitable linear combination of
$\left(U(\beta,\theta)^{+}\right)^{-1}\varphi_1$,
$\varphi_1$ and $U(\beta,\theta)^{+}\varphi_1$ yields $\varphi_3$. Since
$U(\beta,\theta)^{+}\varphi_2=b_1\varphi_1+b_2\varphi_2+b_3\varphi_3+b_4\varphi_4$
we obtain that
$\varphi_4$ can be written as a linear combination desired. Due to the structure of
$U(\beta,\theta)^{+}$,
the process can be iterated to obtain any $\varphi_k$.
\end{proof}

\

We are in conditions to prove pure point spectrum for our model.

\

\begin{proof} {\bf (Theorem~\ref{MainTheorem}(i))} Fix $\beta$ irrational and let
$|\cdot|$ denote the
Lebesgue measure on $[0,2\pi)$. By Theorem~\ref{SpectrumTheorem}{\it (ii)},  for any
$E\in[0,2\pi)$ there exists
$\Omega(E)\subset[0,2\pi)$ with
$|\Omega(E)|=1$ such that
\[\gamma_{\theta}(E)>0,\qquad \forall\;\theta\in\Omega(E).
\] Thus, by Fubini's Theorem,
\begin{eqnarray*}1=\int_0^{2\pi}|\Omega(E)|\frac{dE}{2\pi}&=&
\int_0^{2\pi}\left(\int_0^{2\pi}\chi_{\Omega(E)}(\theta)\frac{d\theta}{2\pi}\right)
\frac{dE}{2\pi}\\ &=&
\int_0^{2\pi}\left(\int_0^{2\pi}\chi_{\Omega(E)}(\theta)\frac{dE}{2\pi}\right)
\frac{d\theta}{2\pi}
\end{eqnarray*} and for $\theta$ in a set of measure one
\[\int_0^{2\pi}\chi_{\Omega(E)}(\theta)\frac{dE}{2\pi}=1,\] that is,
$\theta\in\Omega(E)$ for almost all
$E\in[0,2\pi)$. Then we get the existence of $\Omega_0\subset[0,2\pi)$ with
$|\Omega_0|=1$ such that for
any $\theta\in\Omega_0$ there exists
$A_{\theta}\subset[0,2\pi)$ with $|A_{\theta}|=0$ and
\[\gamma_{\theta}(E)>0,\qquad \forall\;E\in A_{\theta}^{c}:= [0,2\pi)\setminus
A_{\theta}.\] Let
$\mu_{\theta,\lambda}^k$ be the spectral measures associated with
\[ U_{\lambda}(\beta,\theta)^{+}=\int_0^{2\pi}e^{iE}dF_{\theta,\lambda}(E)
\] and respectively vectors $\varphi_k$, so that for $k\in\N^*$ and all Borel sets
$\Lambda\subset[0,2\pi)$
\[\mu_{\theta,\lambda}^k(\Lambda)=\langle\varphi_k,F_{\theta,\lambda}(\Lambda)\varphi_k\rangle.
\] Now, for rank one perturbations of unitary operators there is a spectral
averaging formula as for  rank
one perturbations of self-adjoint operators (see~\cite{S,SW} for the self-adjoint
case and ~\cite{B,C} for
the unitary case). Thus, for any $f\in L^1([0,2\pi))$ one has
\begin{equation}\label{AveragingEquation}
\int_0^{2\pi}d\lambda\int_0^{2\pi}f(E)d\mu_{\theta,\lambda}^1(E)=
\int_0^{2\pi}f(E)\frac{dE}{2\pi}.
\end{equation} Then, applying (\ref{AveragingEquation}) with $f$ the characteristic
function of
$A_{\theta}$ we obtain
\begin{eqnarray*}
0&=&|A_{\theta}|=\int_0^{2\pi}\chi_{A_{\theta}}(E)\frac{dE}{2\pi}\\&=&
\int_0^{2\pi}d\lambda\int_0^{2\pi}\chi_{A_{\theta}}(E)d\mu_{\theta,\lambda}^1(E)=
\int_0^{2\pi}\mu_{\theta,\lambda}^1(A_{\theta})d\lambda,
\end{eqnarray*} and so $\mu_{\theta,\lambda}^1(A_{\theta})=0$ for almost all
$\lambda$. Therefore, for each $\theta\in\Omega_0$, there is
$J_{\theta}\subset[0,2\pi)$ with $|J_{\theta}^{c}|=0$ such that
\begin{equation}\label{CyclicEquation}
\mu_{\theta,\lambda}^1(A_{\theta})=0,\qquad\forall\;\lambda\in J_{\theta}.
\end{equation} By Lemma~\ref{CyclicLemma} and (\ref{CyclicEquation}), it follows that
$F_{\theta,\lambda}(A_{\theta})=0$ for all $\theta\in\Omega_0$ and
$\lambda\in J_{\theta}$. Moreover, let
$ S_{\theta,\lambda}$ denote the set of $E\in[0,2\pi)$ so that the equation
\[ U_{\lambda}(\beta,\theta)^{+}\psi=e^{iE}\psi
\] has a nontrivial polynomially bounded solution. It is known that
\[ F_{\theta,\lambda}\left([0,2\pi)\setminus S_{\theta,\lambda}\right)=0
\] (see~\cite{BHJ,HJS}). Thus we conclude that $S_{\theta,\lambda}\cap
A_{\theta}^{c}$ is a support for
$F_{\theta,\lambda}(\cdot)$ (see remark bellow) for all
$\theta\in\Omega_0$ and $\lambda\in J_{\theta}$.

Now, if $E\in S_{\theta,\lambda}\cap A_{\theta}^{c}$ then
$U_{\lambda}(\beta,\theta)^{+}\psi=e^{iE}\psi$ has a nontrivial polynomially bounded
solution $\psi$ and
$\gamma_{\theta}(E)>0$. By construction
$\gamma_{\theta,\lambda}(E)=\gamma_{\theta}(E)$ where
$\gamma_{\theta,\lambda}(E)$ is the Lyapunov exponent associated with
$U_{\lambda}(\beta,\theta)^{+}$. Thus, by Oseledec's Theorem, every solution which
is polynomially bounded
necessarily has to decay exponentially, so $\psi$ is in
$l^2(\N^*)$ and is an eigenfunction of
$U_{\lambda}(\beta,\theta)^{+}$. Hence, we conclude that each
$E\in S_{\theta,\lambda}\cap A_{\theta}^{c}$ is an eigenvalue of
$U_{\lambda}(\beta,\theta)^{+}$ with corresponding eigenfunction decaying
exponentially. As $l^2(\N^*)$ is
separable, it follows that $S_{\theta,\lambda}\cap A_{\theta}^{c}$ is countable and
then
$F_{\theta,\lambda}(\cdot)$ has countable support for all
$\theta\in\Omega_0$ and $\lambda\in J_{\theta}$. Thus
$U_{\lambda}(\beta,\theta)^{+}$ has purely point spectrum for a.e.\
$\theta,\;\lambda\in[0,2\pi)$.
\end{proof}

\

\noindent \textit{Remark.} We say that a Borel set $S$ supports the spectral
projection $F(\cdot)$ if
$F([0,2\pi)\setminus S)=0$.

\

\section{Energy Instability}
\label{InstabilitySection} In this section we present the proof of
Theorem~\ref{MainTheorem}{\it (ii)}. The
initial strategy is that of Appendix~2 of~\cite{dRJLS}, and  Lemmas~\ref{Lemma1}
and~\ref{Lemma2} ahead are
 similar to Lemmas B.1 and B.2 in
\cite{dRJLS}. However, some important technical issues
needed quite different arguments. To begin with we shall discuss a series of
preliminary lemmas, adapted to
the unitary case from the self-adjoint setting.

\

\subsection{Preliminary Lemmas} Let $P_{n\geq a}$ denote the projection onto those
vectors supported by
$\{n:n\geq a\}$, that is, for $\psi\in l^2(\N^*)$
\[(P_{n\geq a}\psi)(n)=\left\{\begin{array}{ccc} 0, & \mbox{if} & n< a\\
\psi(n), & \mbox{if} & n\geq a
\end{array}\right.,\] and similarly for
$P_{n< a}$. Let $f(n)$ be a monotone increasing sequence with
$f(n)\rightarrow\infty$ as $n\to\infty$.

\begin{Lemma}\label{Lemma1} If there exists $T_m\rightarrow\infty$,
$T_m\in\N$ for all $m$, such that
\begin{equation}\label{Lemma1Equation}
\frac{1}{T_m+1}\sum_{j=T_m}^{2T_m}\|P_{n\geq\frac{T_m}{f(T_m)}}
\left(U_{\lambda}(\beta,\theta)^{+}\right)^j\varphi_1\|^2\geq\frac{1}{f(T_m)^2}\;,
\end{equation} then
\[\limsup_{j\rightarrow\infty}\|X\left(U_{\lambda}(\beta,\theta)^{+}\right)^j
\varphi_1\|^2\frac{f(j)^5}{j^2}=\infty.
\]
\end{Lemma}
\begin{proof} By hypothesis, for each $m\in\N$, there must be some
$j_m\in[T_m,2T_m]$ such that \[\|P_{n\geq\frac{T_m}{f(T_m)}}
\left(U_{\lambda}(\beta,\theta)^{+}\right)^{j_m}\varphi_1\|^2\geq\frac{1}{f(T_m)^2}\]
and then
\begin{eqnarray*}
\|X\left(U_{\lambda}(\beta,\theta)^{+}\right)^{j_m}\varphi_1\|^2 &=&
\sum_{n\in\N^*}n^2|\left(U_{\lambda}(\beta,\theta)^{+}\right)^{j_m}\varphi_1(n)|^2\\
&\geq&
\sum_{n\in\N^*}\Big|\frac{T_m}{f(T_m)}\left(P_{n\geq\frac{T_m}{f(T_m)}}
\left(U_{\lambda}(\beta,\theta)^{+}\right)^{j_m}\varphi_1\right)(n)\Big|^2\\
&\geq&\frac{T_m^2}{f(T_m)^4}.
\end{eqnarray*} Therefore
\begin{eqnarray*}
\frac{f(j_m)^5}{j_m^2}\|X\left(U_{\lambda}(\beta,\theta)^{+}\right)^{j_m}\varphi_1\|^2
&\geq&
\left(\frac{T_m}{j_m}\right)^2\left(\frac{f(j_m)}{f(T_m)}\right)^4f(j_m)\\ &\geq&
\frac{1}{4}f(j_m)\rightarrow\infty
\end{eqnarray*} and the lemma is proved.
\end{proof}

\

In order to prove Theorem~\ref{MainTheorem}{\it (ii)} we want to apply the above
lemma with
$f(n)=(\ln(n+2))^{1/5}$. By keeping this goal in mind, the estimate in relation
(\ref{Lemma1Equation}) is
crucial as well as the following lemmas. 

\

\begin{Lemma}\label{Lemma2} Let $\xi$ be a unit vector, $P$ a projection, and $U$ a
unitary operator. If
$\xi=\eta+\psi$ with $\langle\eta,\psi\rangle=0$, then
\begin{equation}\label{Lemma2Equation}
\frac{1}{T+1}\sum_{j=T}^{2T}\|(\Id-P)U^j\xi\|^2\geq\|\psi\|^2-3\left(
\frac{1}{T+1}\sum_{j=T}^{2T}\|PU^j\psi\|^2\right)^{1/2}.
\end{equation}
\end{Lemma}
\begin{proof} Denote
$D:=\frac{1}{T+1}\sum_{j=T}^{2T}\|(\Id-P)U^j\xi\|^2$. Then
\begin{eqnarray*} D &=& \frac{1}{T+1}\sum_{j=T}^{2T}(1-\|PU^j\xi\|^2)\\ &=&
\frac{1}{T+1}\sum_{j=T}^{2T}(\|\psi\|^2+\|\eta\|^2-\|PU^j(\eta+\psi)\|^2)\\ &=&
\|\eta\|^2-\frac{1}{T+1}\sum_{j=T}^{2T}\|PU^j\eta\|^2\\ & & +
\|\psi\|^2-\frac{1}{T+1}\sum_{j=T}^{2T}(\|PU^j\psi\|^2 +2\Rem(\langle
PU^j\eta,PU^j\psi\rangle))\\ &=& A+B,
\end{eqnarray*} with
$A=\|\eta\|^2-\frac{1}{T+1}\sum_{j=T}^{2T}\|PU^j\eta\|^2$ and
$ B=\|\psi\|^2-\frac{1}{T+1}\sum_{j=T}^{2T}(\|PU^j\psi\|^2 \break 
+2\Rem(\langle PU^j\eta,PU^j\psi\rangle)).
$ 

Clearly,
$\frac{1}{T+1}\sum_{j=T}^{2T}\|PU^j\eta\|^2\leq
\|\eta\|^2\leq1,
$ and the same is true with $\eta$ replaced by $\psi$. Hence
$A\geq0$ and
\begin{eqnarray*} B &=& \|\psi\|^2-\frac{1}{T+1}\sum_{j=T}^{2T}\|PU^j\psi\|^2
-\frac{2}{T+1}\sum_{j=T}^{2T}\Rem(\langle PU^j\eta,PU^j\psi\rangle)\\ &\geq&
\|\psi\|^2-\frac{1}{T+1}\sum_{j=T}^{2T}\|PU^j\psi\|^2
-\frac{2}{T+1}\sum_{j=T}^{2T}\|PU^j\eta\|\|PU^j\psi\|\\ &\geq &
\|\psi\|^2-\Big(\frac{1}{T+1}\sum_{j=T}^{2T}\|PU^j\psi\|^2\Big)^{\frac{1}{2}}\\ & &
-2\Big(\frac{1}{T+1}\sum_{j=T}^{2T}\|PU^j\eta\|^2\Big)^{\frac{1}{2}}
\Big(\frac{1}{T+1}\sum_{j=T}^{2T}\|PU^j\psi\|^2\Big)^{\frac{1}{2}}\\ &\geq&
\|\psi\|^2-3\Big(\frac{1}{T+1}\sum_{j=T}^{2T}\|PU^j\psi\|^2\Big)^{\frac{1}{2}}.
\end{eqnarray*} The result follows immediately.
\end{proof}

\

The following lemma is an adaptation to the discrete setup of a classical estimate
found in Lemma~4.5, page
543 of \cite{Kato}.

\begin{Lemma}\label{Lemma3} Let $U=\int_0^{2\pi}e^{it}dE_U(t)$ be the spectral
decomposition of a unitary
operator $U$ on the Hilbert space
$\mathcal{H}$. Let $\xi\in\mathcal{H}$ be an absolutely continuous vector for $U$,
i.e., the spectral
measure $\mu_{\xi}$, associated to $U$ and $\xi$, is absolutely continuous with
respect to Lebesgue
measure, and denote by $g=\frac{d\mu_{\xi}}{dx}\in L^1([0,2\pi))$ the corresponding
Radon-Nikodym
derivative. Define
\[\trip{\xi}_U=\left\|g\right\|_{\infty}^{1/2}.\] Then, for any
$\eta\in\mathcal{H}$, one has
\[\sum_{j\in\Z}|\langle U^j\xi,\eta\rangle|^2\leq2\pi\,\trip{\xi}^2_U\,\|\eta\|^2.\]
If it is clear the unitary operator in question, then $\trip{\cdot}$ will be used to
indicate
$\trip{\cdot}_U$.
\end{Lemma}
\begin{proof} If $\trip{\xi}=\infty$ then the result is clear. Suppose
$\trip{\xi}<\infty$ and take
$\eta\in\mathcal{H}$. Denote by $P_{ac}$ the spectral projection onto the absolutely
continuous subspace
$\mathcal{H}_{ac}$ of
$U$, $\eta_0=P_{ac}\eta$ and
$\tilde{g}=\frac{d\mu_{\eta_0}}{d\lambda}$; then  $\mu_{\xi,\eta}$ is absolutely
continuous and its
Radon-Nikodym derivative $h$ is estimate by
\[|h(x)|\leq(g\tilde{g})^{\frac{1}{2}}(x)=g^{\frac{1}{2}}(x)\,\tilde{g}^{\frac{1}{2}}(x)
\leq\trip{\xi}\;\tilde{g}^{\frac{1}{2}}(x).
\] Hence $h\in L^2([0,2\pi))$ with $L^2$ norm estimated by
\[\|h\|_2\leq\trip{\xi}\Big(\int_0^{2\pi}\tilde{g}(x)
dx\Big)^{\frac{1}{2}}=\trip{\xi}\Big(\int_0^{2\pi}
d\mu_{\eta_0}\Big)^{\frac{1}{2}}
\]
\[=\trip{\xi}\cdot\|\eta_0\|\leq\trip{\xi}\cdot\|\eta\|.
\] Since
 \[\langle U^j\xi,\eta\rangle=\int_0^{2\pi}e^{ijt}d\mu_{\xi,\eta}(t)=
\int_0^{2\pi}e^{ijt}h(t)dt=\sqrt{2\pi}(\mathcal{F}h)(j),\] it follows that
\[\sum_{j\in\Z}|\langle U^j\xi,\eta\rangle|^2=\sum_{j\in\Z}2\pi|(\mathcal{F}h)(j)|^2=
2\pi\|h\|_2^2\leq2\pi\trip{\xi}^2\|\eta\|^2,\] which is precisely the stated result.
\end{proof}

\

\subsection{Cauchy and Borel Transforms} Given a probability measure $\mu$ on
$\partial\mathbb{D}=\{z\in\C:|z|=1\}$, its Cauchy $F_{\mu}(z)$ and Borel
$R_{\mu}(z)$ transforms are, respectively, for
$z\in\C$ with
$|z|\neq1$,
\[F_{\mu}(z)=\int_{\partial\mathbb{D}}\frac{e^{it}+z}{e^{it}-z}d\mu(t)\] and
\[R_{\mu}(z)=\int_{\partial\mathbb{D}}\frac{1}{e^{it}-z}d\mu(t).\]
$R_{\mu}$ is related to $F_{\mu}$ by
\begin{equation}\label{RelationEquation} F_{\mu}(z)=2zR_{\mu}(z)+1.
\end{equation} Moreover, $F_{\mu}$ has the following properties~\cite{S1}:
\begin{itemize}
\item $\lim_{r\uparrow1}F_{\mu}(re^{i\theta})$ exists for a.e.\ $\theta$, and if one
decomposes the measure
in its absolutely continuous and singular parts
\[d\mu(\theta)=\omega(\theta)\frac{d\theta}{2\pi}+d\mu_s(\theta),\] then
\begin{equation}\label{MeasureEquation}
\omega(\theta)= \lim_{r\uparrow1}\Rem F_{\mu}(re^{i\theta}).
\end{equation}
\item $\theta_0$ is a pure point of $\mu$ if and only if
$\lim_{r\uparrow1}(1-r)\Rem F_{\mu}(re^{i\theta_0})\neq0$.
\item $d\mu_s$ is supported on
$\{\theta:\lim_{r\uparrow1}F_{\mu}(re^{i\theta})=\infty\}$.
\end{itemize}

\

Now, let $U$ be a unitary operator on a separable Hilbert space
$\mathcal{H}$ and $\phi$ a cyclic vector for $U$. Consider the rank one perturbation
of $U$
\[ U_{\lambda}=Ue^{i\lambda P_{\phi}}=U(\Id+(e^{i\lambda}-1)P_{\phi}),
\] where
$P_{\phi}(\cdot)=\langle\phi,\cdot\rangle\phi$ and
$\lambda\in[0,2\pi)$. Denote by $d\mu_{\lambda}$ the spectral measure associated
with $U_{\lambda}$ and
$\phi$,
$F_{\lambda}=F_{\mu_{\lambda}}$ and
$R_{\lambda}=R_{\mu_{\lambda}}$. We have the following relations between
$R_{\lambda}$ and $R_0$,
$F_{\lambda}$ and $F_0$:

\

\begin{Lemma}\label{Relations} For $|z|\neq1$
\begin{equation}\label{BorelEquation} R_{\lambda}(z)=
\frac{R_0(z)}{e^{i\lambda}+z(e^{i\lambda}-1)R_0(z)}
\end{equation} and
\begin{equation}\label{CauchyEquation}
F_{\lambda}(z)=\frac{(e^{i\lambda}-1)+(e^{i\lambda}+1)F_0(z)}
{(e^{i\lambda}+1)+(e^{i\lambda}-1)F_0(z)}
\end{equation} In particular, for $\lambda\neq\pi$,
\begin{equation}\label{RealEquation}
\Rem F_{\lambda}(z)=\frac{(1+y^2)\Rem F_0(z)} {|1+iyF_0(z)|^2},
\end{equation} where $y=\frac{\sin\lambda}{1+\cos\lambda}$, and for
$\lambda=\pi$
\begin{equation}\label{RealEquation1}
\Rem F_{\lambda}(z)=\frac{\Rem F_0(z)} {|F_0(z)|^2}.
\end{equation}
\end{Lemma}
\begin{proof} Relation (\ref{BorelEquation}) was got in~\cite{C}. For checking
(\ref{CauchyEquation}) we
use  relations (\ref{RelationEquation}) and (\ref{BorelEquation}). In fact,
\begin{eqnarray*} F_{\lambda}(z) &=& 2zR_{\lambda}(z)+1 \\ &=& 2z
\frac{R_0(z)}{e^{i\lambda}+z(e^{i\lambda}-1)R_0(z)}+1\\ &=&
\frac{e^{i\lambda}+z(e^{i\lambda}-1)R_0(z)+2zR_0(z)}{e^{i\lambda}+z(e^{i\lambda}-1)R_0(z)}\\
&=&
\frac{e^{i\lambda}+z(e^{i\lambda}+1)R_0(z)}{e^{i\lambda}+z(e^{i\lambda}-1)R_0(z)}\\ &=&
\frac{2e^{i\lambda}+2ze^{i\lambda}R_0(z)+2zR_0(z)}
{2e^{i\lambda}+2ze^{i\lambda}R_0(z)-2zR_0(z)}\\ &=&
\frac{e^{i\lambda}-1+e^{i\lambda}+2e^{i\lambda}zR_0(z)+1+2zR_0(z)}
{e^{i\lambda}+1+e^{i\lambda}+2e^{i\lambda}zR_0(z)-1-2zR_0(z)}\\ &=&
\frac{(e^{i\lambda}-1)+(e^{i\lambda}+1)(1+2zR_0(z))}
{(e^{i\lambda}+1)+(e^{i\lambda}-1)(1+2zR_0(z))}\\ &=&
\frac{(e^{i\lambda}-1)+(e^{i\lambda}+1)F_0(z)}
{(e^{i\lambda}+1)+(e^{i\lambda}-1)F_0(z)}.
\end{eqnarray*} Now, for $\lambda\neq\pi$ we have $e^{i\lambda}+1\neq0$ and then
\begin{eqnarray*} F_{\lambda}(z) &=&
\frac{\frac{e^{i\lambda}-1}{e^{i\lambda}+1}+F_0(z)}
{1+\left(\frac{e^{i\lambda}-1}{e^{i\lambda}+1}\right)F_0(z)}\\ &=&
\frac{iy+F_0(z)}{1+iyF_0(z)}
\times\frac{1-iy\overline{F_0(z)}}{1-iy\overline{F_0(z)}}\\ &=&
\frac{iy+F_0(z)-iy|F_0(z)|^2+y^2\overline{F_0(z)}}{|1+iyF_0(z)|^2},
\end{eqnarray*} where $\frac{e^{i\lambda}-1}{e^{i\lambda}+1}=iy$ and
$y=\frac{\sin\lambda}{1+\cos\lambda}$. So, for $\lambda\neq\pi$,
\[\Rem F_{\lambda}(z)=\frac{(1+y^2)\Rem F_0(z)}{|1+iyF_0(z)|^2}\] and
(\ref{RealEquation}) is obtained. For
$\lambda=\pi$ we have $F_{\lambda}(z)=\frac{1}{F_0(z)}$ and (\ref{RealEquation1})
follows.
\end{proof}

\

\begin{Lemma}\label{Lemma4} Fix a rational number $\beta$. Then there exist $C_1>0$ and
$C_2<\infty$, and for each $\theta\in[0,2\pi)$ and
$\lambda\in[\frac{\pi}{6},\frac{\pi}{2}]$ a decomposition
\[\varphi_1=\eta_{\theta,\lambda}+\psi_{\theta,\lambda}\] so that
\begin{equation}\label{Lemma4Equation}
\langle\eta_{\theta,\lambda},\psi_{\theta,\lambda}\rangle=0,
\end{equation}
\begin{equation}\label{Lemma4Equation1}
\|\psi_{\theta,\lambda}\|\geq C_1,
\end{equation}
\begin{equation}\label{Lemma4Equation2}
\trip{\psi_{\theta,\lambda}}_{U_{\lambda}(\beta,\theta)^{+}}\leq C_2
\end{equation}
(the notation $\trip{\cdot}_U$ was introduced in Lemma~\ref{Lemma3}).
\end{Lemma}
\begin{proof} We break the proof in some steps.

\underline{Step 1}.  By Theorem~\ref{SpectrumTheorem}, since
$\beta$ is rational,
\[
\sigma_{\mathrm{sc}}(U(\beta,\theta)^{+})=\emptyset,\quad\sigma_{\mathrm{ac}}(U(\beta,\theta)^{+})=
\sigma_{\mathrm{ac}}(U(\beta,\theta))
\] and the point spectrum of
$U(\beta,\theta)^{+}$ consists of finitely many simple eigenvalues in the resolvent
set of
$U(\beta,\theta)$. Denote by $\mu_{\theta,\lambda}$ the spectral measure associated to
$U_{\lambda}(\beta,\theta)^{+}$ and (the cyclic vector) $\varphi_1$, and by
$\mu_{\theta}$ the spectral
measure associated to
$U(\beta,\theta)^{+}$ and
$\varphi_1$ (i.e., the case $\lambda=0$). Write
\[d\mu_{\theta,\lambda}(E)=f_{\theta,\lambda}(E)\frac{dE}{2\pi}+
d\mu^{\theta,\lambda}_s(E),
\]
\[d\mu_{\theta}(E)=f_{\theta}(E)\frac{dE}{2\pi}+d\mu^{\theta}_p(E).
\]

\

\underline{Step 2}. Relation between $f_{\theta,\lambda}$ and
$f_{\theta}$: By Lemma~\ref{Relations}, for
$\lambda\neq\pi$ one has
\[\Rem F_{\mu_{\theta,\lambda}}(z)=\frac{(1+y^2)\Rem F_{\mu_{\theta}}(z)}
{\left|1+iyF_{\mu_{\theta}}(z)\right|^2},
\] where $y=\frac{\sin\lambda}{1+\cos\lambda}$ and then
\[f_{\theta,\lambda}(E)=\frac{(1+y^2)f_{\theta}(E)}
{\left|1+iy\lim_{r\uparrow1}F_{\mu_{\theta}}(re^{iE})\right|^2},\] for almost all $E$.

\

\underline{Step 3}. Relation between $f_{\theta}$ and $f_0$: By
(\ref{ThetaEquation}) and
(\ref{PositiveEquation}) one gets
\begin{equation}\label{Step3Equation} U(\beta,\theta)^{+}=e^{-i2\theta}U(\beta,0)^{+}
\end{equation} and using this relation it found that
\[\left(U(\beta,\theta)^{+}\right)^j=e^{-ij2\theta}\left(U(\beta,0)^{+}\right)^j\]
for all $j\in\Z$. Thus,
by the spectral theorem, for any $j\in\Z$,
\[\int_0^{2\pi}e^{-ijE}f_{\theta}(E)\frac{dE}{2\pi}=
\int_0^{2\pi}e^{-ijE}f_0(E-2\theta)\frac{dE}{2\pi}.
\] Hence
\begin{equation}\label{Lemma4Equation3} f_{\theta}(E)=f_0(E-2\theta)
\end{equation} for almost all $E$.

\

\underline{Step 4}. Lower and upper bounds for
$f_{\theta,\lambda}$: We have \[\lim_{r\uparrow1} F_{\mu_{\theta}}(re^{iE})=
f_{\theta}(E)+i\lim_{r\uparrow1}\Imm F_{\mu_{\theta}}(re^{iE})\] and
\begin{eqnarray*}
\lim_{r\uparrow1}\Imm F_{\mu_{\theta}}(re^{iE})&=&
\lim_{r\uparrow1}\int_0^{2\pi}\Imm\left(\frac{e^{it}+re^{iE}}{e^{it}-re^{iE}}\right)
f_{\theta}(t)\frac{dt}{2\pi}\\ &&
+\lim_{r\uparrow1}\int_0^{2\pi}\Imm\left(\frac{e^{it}+re^{iE}}{e^{it}-re^{iE}}\right)
d\mu^{\theta}_p(t).
\end{eqnarray*}  If we denote
\[ g_{\theta}(E)= \lim_{r\uparrow1}\int_0^{2\pi}\Imm
\left(\frac{e^{it}+re^{iE}}{e^{it}-re^{iE}}\right) f_{\theta}(t)\frac{dt}{2\pi},
\] then by (\ref{Lemma4Equation3}) we obtain $g_{\theta}(E)=g_0(E-2\theta)$ for
almost all $E$. On the
other hand, by (\ref{Step3Equation}) we have that $E$ is an eigenvalue of
$U(\beta,\theta)^{+}$ if and only
if $E-2\theta$ is an eigenvalue of $U(\beta,0)^{+}$. Let
$\{E_j^{\theta}\}_{j=1}^n$ be the set of eigenvalues of
$U(\beta,\theta)^{+}$ (recall that $n<\infty$) and
$d\mu_p^{\theta}=\sum_{j=1}^n\kappa_j^{\theta}\delta_{E_j^{\theta}}$ ($\delta_E$ is
the Dirac measure at
$E$). Then
\begin{eqnarray*}
\lim_{r\uparrow1}\int_0^{2\pi}\Imm\left(\frac{e^{it}+re^{iE}}{e^{it}-re^{iE}}\right)
d\mu^{\theta}_p(t)&=&
\lim_{r\uparrow1}\int_0^{2\pi}\frac{2r\sin(E-t)}{1+r^2-2r\cos(E-t)}d\mu^{\theta}_p(t)\\
&=&\lim_{r\uparrow1}\sum_{j=1}^n\frac{2r\sin(E-E_j^{\theta})\kappa_j^{\theta}}
{1+r^2-2r\cos(E-E_j^{\theta})}\\ &=&
\sum_{j=1}^n\frac{2\sin(E-2\theta-E_j^0)\kappa_j^{\theta}}
{\left|e^{iE_j^0}-e^{i(E-2\theta)}\right|^2}.
\end{eqnarray*} Since $f_0\in L^1([0,2\pi))$, by a result of~\cite{K} (Theorem 1.6
in Chapter III), the
function $g_0$ is of weak $L^1$ type, i.e., $g_0$ is measurable  and there exits a
constant $C>0$ such that
for all
$T>0$ the Lebesgue measure
\begin{equation}\label{WeakEquation}
\left\vert\{E:\vert g_0(E)\right\vert\leq T\}\vert\geq1-\frac{C}{T}.
\end{equation}  Pick $S>0$ such that
$\Omega_S:=\Big\{E:\frac{1}{S}\leq f_0(E)\leq S\Big\}$ satisfies
$\vert\Omega_S\vert>0$ and
$\text{dist}\;(\Omega_S,\{E_j^0\}_{j=1}^n)=L>0$. Then choose $T$ sufficiently large
such that
\[A:=\Omega_S\cap\{E:\vert g_0(E)\vert\leq T\}\] satisfies $\vert A\vert>0$; by
(\ref{WeakEquation}) this
is possible. For $\theta\in[0,2\pi)$ put
\[I_{\theta}:=\{E\in[0,2\pi):E-2\theta\in A\};
\] thus $\vert I_{\theta}\vert=\vert A\vert>0$. Then, for all $\theta\in[0,2\pi)$,
$\lambda\in[0,\frac{\pi}{2}]$ (equivalently
$y\in[0,1]$) and almost all $E\in I_{\theta}$ one has
\begin{eqnarray*}
\left\vert1+iy\lim_{r\uparrow1} F_{\mu_{\theta}}(re^{iE})\right\vert &\leq& 1+\vert
y\vert\Bigg(
f_{\theta}(E)+\vert g_{\theta}(E)\vert\\ && +
\left|\sum_{j=1}^n\frac{2\sin(E-2\theta-E_j^0)\kappa_j^{\theta}}
{|e^{iE_j^0}-e^{i(E-2\theta)}|^2}\right|\Bigg)\\ &\leq& 1+ f_0(E-2\theta)+\vert
g_0(E-2\theta)\vert+
\sum_{j=1}^n\frac{2|\kappa_j^{\theta}|}{L^2}\\ &\leq& 1+S+T+\frac{2}{L^2}.
\end{eqnarray*} So, for all $\theta\in[0,2\pi)$,
$\lambda\in[0,\frac{\pi}{2}]$ and almost all $E\in I_{\theta}$
\begin{eqnarray*}
 f_{\theta,\lambda}(E) &=& \frac{(1+y^2)f_{\theta}(E)}
{\vert1+iy\lim_{r\uparrow1}
F_{\mu_{\theta}}(re^{iE}))\vert^2}\\ &\geq&
\frac{f_0(E-2\theta)}{(1+S+T+2/L^2)^2}\\ &\geq&
\frac{1}{S(1+S+T+2/L^2)^2}.
\end{eqnarray*}

In order to get un upper  bound, note that
\[
\left\vert1+iy\lim_{r\uparrow1} F_{\mu_{\theta}}(re^{iE})\right\vert\geq
yf_{\theta}(E)\ge0,\] and so, for all
$\theta\in[0,2\pi)$,
$\lambda\in[\frac{\pi}{6},\frac{\pi}{2}]$ (equivalently
$y\in\Big[\frac{1}{2+\sqrt{3}},1\Big]$) and almost all $E\in I_{\theta}$
\begin{eqnarray*}
 f_{\theta,\lambda}(E) &=&
\frac{(1+y^2)f_{\theta}(E)}{\left\vert1+iy\lim_{r\uparrow1}
F_{\mu_{\theta}}(re^{iE})\right\vert^2}\\ &\leq&
\frac{(1+y^2)f_{\theta}(E)}{y^2f_{\theta}(E)^2} =\frac{(1+y^2)}{y^2f_0(E-2\theta)}\\
&\leq&
2(2+\sqrt{3})^2S.
\end{eqnarray*}

Summing up, for all $\theta\in[0,2\pi)$,
$\lambda\in[\frac{\pi}{6},\frac{\pi}{2}]$ and almost all $E\in I_{\theta}$, we have
proved that
\begin{equation}\label{Lemma4Equation4}
\frac{1}{S(1+S+T+2/L^2)^2}\leq
f_{\theta,\lambda}(E)\leq2(2+\sqrt{3})^2S.
\end{equation}

\

\underline{Step 5}. Conclusion: For
$\lambda\in[\frac{\pi}{6},\frac{\pi}{2}]$ and $\theta\in[0,2\pi)$ let
\[
\psi_{\theta,\lambda}=P_{I_{\theta}}^{\theta,\lambda}\varphi_1,\quad
\eta_{\theta,\lambda}=(\Id-P_{I_{\theta}}^{\theta,\lambda})\varphi_1,
\] where
$P_{I_{\theta}}^{\theta,\lambda}$ is the spectral projection (of
$U_{\lambda}(\beta,\theta)^{+}$) onto
$I_{\theta}$. Then for each $\theta\in[0,2\pi)$ and
$\lambda\in[\frac{\pi}{6},\frac{\pi}{2}]$ we have the decomposition
$\varphi_1=\psi_{\theta,\lambda}+\eta_{\theta,\lambda}$ that satisfies
(\ref{Lemma4Equation}). 

By the construction in Step 4, we have that $A=I_0$ is in the
absolutely continuous spectrum of $U(\beta,0)^{+}$, so by (\ref{Step3Equation}) and
the definition of $I_{\theta}$ it follows that $I_{\theta}$ is in the absolutely
continuous spectrum of $U(\beta,\theta)^{+}$; thus using Birman-Krein's theorem
on invariance of absolutely continuous spectrum under trace class perturbations, we
conclude that $I_{\theta}$ belongs to the absolutely continuous spectrum of
$U_{\lambda}(\beta,\theta)^{+}$ for all $\lambda$. Therefore by
(\ref{Lemma4Equation4})
\begin{eqnarray*}
\Vert\psi_{\theta,\lambda}\Vert^2 &=&
\langle\psi_{\theta,\lambda},\psi_{\theta,\lambda}\rangle =
\langle
P_{I_{\theta}}^{\theta,\lambda}\varphi_1,P_{I_{\theta}}^{\theta,\lambda}\varphi_1
\rangle\\ &=&
\langle\varphi_1,P_{I_{\theta}}^{\theta,\lambda}\varphi_1\rangle =
\int_0^{2\pi}\chi_{I_{\theta}}(E)d\mu_{\theta,\lambda}\\ &=&
 \int_{I_{\theta}}f_{\theta,\lambda}(E)\frac{dE}{2\pi} \geq
 \frac{\vert A\vert}{2\pi S(1+S+T+2/L^2)^2}
\end{eqnarray*} and (\ref{Lemma4Equation1}) holds with
\[ C_1=\Big(\frac{\vert A\vert}{2\pi S(1+S+T+2/L^2)^2}
\Big)^{1/2}>0;
\] also
\[\trip{\psi_{\theta,\lambda}}_{U_{\lambda}(\beta,\theta)^{+}}^{2}=
\trip{P_{I_{\theta}}^{\theta,\lambda}\varphi_1}_{U_{\lambda}(\beta,\theta)^{+}}^{2}=
\Vert\chi_{I_{\theta}}f_{\theta,\lambda}\Vert_{\infty}\leq2(2+\sqrt{3})^{2}S\] and
(\ref{Lemma4Equation2})
holds with
$C_2=(2(2+\sqrt{3})^{2}S)^{1/2}<\infty$. The lemma is proved.
\end{proof}

\

\subsection{Variation of $\beta$} The next lemma gives an estimate of the dependence
of the dynamics on
$\beta$. Its proof strongly uses
 the structure of $U_{\lambda}(\beta,\theta)^{+}$.

\

\begin{Lemma}\label{Lemma5} Let $\beta,\beta'\in\R$. Then, for $n\geq1$,
\[
\left\|\left(U_{\lambda}(\beta,\theta)^{+}\right)^n\varphi_1-
\left(U_{\lambda}(\beta',\theta)^{+}\right)^n\varphi_1\right\|
\leq2\times4^n(2n^2-n)2\pi\left|\beta-\beta'\right|.\]
\end{Lemma}
\begin{proof} It is an induction. We have
\begin{eqnarray*} U_{\lambda}(\beta,\theta)^{+}\varphi_j &=&
U(\beta,\theta)^{+}(\Id+(e^{i\lambda}-1)
P_{\varphi_1})\varphi_j\\ &=& \left\{\begin{array}{ccc} U(\beta,\theta)^{+}\varphi_j
& \mbox{if} & j>1\\
U(\beta,\theta)^{+}\varphi_1+(e^{i\lambda}-1)U(\beta,\theta)^{+}\varphi_1 &
\mbox{if} & j=1
\end{array}\right.\\ &=& \left\{\begin{array}{ccc} U(\beta,\theta)^{+}\varphi_j &
\mbox{if} & j>1\\
e^{i\lambda}U(\beta,\theta)^{+}\varphi_1 & \mbox{if} & j=1
\end{array}\right.
\end{eqnarray*} Thus
\[U_{\lambda}(\beta,\theta)^{+}\varphi_1= e^{i\lambda}U(\beta,\theta)^{+}\varphi_1=
a_1e^{-i(2\pi\beta)}\varphi_{1}+a_2e^{-i(2\pi\beta)}\varphi_{2}\] where
$a_1=re^{i\lambda}e^{-i(\alpha+2\theta)}$ and
$a_2=ite^{i\lambda}e^{-i(\alpha+2\theta)}$. Since
\begin{equation}\label{StarEquation} |e^{-ix}-e^{-ix'}|\leq2|x-x'|
\end{equation} and $|a_j|\leq1$, $j=1,2$, then
\begin{eqnarray*}
\left\|U_{\lambda}(\beta,\theta)^{+}\varphi_1-U_{\lambda}(\beta',\theta)^{+}\varphi_1\right\|
&\leq&
2\left|e^{-i(2\pi\beta)}-e^{-i(2\pi\beta')}\right|\\ &\leq &
4\times2|2\pi\beta-2\pi\beta'|=2\times4\times2\pi\left|\beta-\beta'\right|
\end{eqnarray*} and the lemma is proved for $n=1$.

Now
\begin{eqnarray*} \left(U_{\lambda}(\beta,\theta)^{+}\right)^2\varphi_1 &=&
U_{\lambda}(\beta,\theta)^{+}U_{\lambda}(\beta,\theta)^{+}\varphi_1\\ &=&
U_{\lambda}(\beta,\theta)^{+}(a_1e^{-i(2\pi\beta)}
\varphi_{1}+a_2e^{-i(2\pi\beta)}\varphi_{2})\\ &=&
e^{i\lambda}a_1e^{-i(2\pi\beta)}U(\beta,\theta)^{+}\varphi_{1}
+a_2e^{-i(2\pi\beta)}U(\beta,\theta)^{+}\varphi_{2}\\ &=&
e^{i\lambda}a_1e^{-i(2\pi\beta)}
\Big(b_1e^{-i(2\pi\beta)}\varphi_{1} +b_2e^{-i(2\pi\beta)}\varphi_{2}\Big)\\ & &
+a_2e^{-i(2\pi\beta)}\Big(c_1e^{-i(3.(2\pi\beta))}\varphi_1
+c_2e^{-i(3.(2\pi\beta))}\varphi_2\\ & &
+c_3e^{-i(5.(2\pi\beta))}\varphi_3 +c_4e^{-i(5.(2\pi\beta))}\varphi_4\Big)
\end{eqnarray*} Since $|a_j|<1$, $|b_j|<1$, $|c_j|<1$ and there are
$2+4<4\times4$ terms in the expansion of
$\left(U_{\lambda}(\beta,\theta)^{+}\right)^2\varphi_1$ and the largest exponent
(which provides the
largest contribution by (\ref{StarEquation})) is obtained from the product of the
exponentials
$e^{-i(2\pi\beta)}e^{-i((2+3)2\pi\beta)}=e^{-i((1+2+3)2\pi\beta)}$, we obtain 
\begin{eqnarray*}
\left\|\left(U_{\lambda}(\beta,\theta)^{+}\right)^2\varphi_1-
\left(U_{\lambda}(\beta',\theta)^{+}\right)^2\varphi_1\right\|
&\leq&4\times4\times2(1+2+3)2\pi\left|\beta-\beta'\right|\\&=&2\times4^2(1+2+3)2\pi\left|\beta-\beta'\right|,
\end{eqnarray*} and the lemma is proved for $n=2$. In a similar way by the structure of
$U_{\lambda}(\beta,\theta)^{+}$ we conclude that
$\left(U_{\lambda}(\beta,\theta)^{+}\right)^3\varphi_1$ has at  most $4^2\times4$
terms where the largest
exponent is in
$e^{-i(1+2+3)2\pi\beta}e^{-i((4+5)2\pi\beta)}=e^{-i((1+2+3+4+5)2\pi\beta)}$ and so
\begin{eqnarray*} & &
\left\|\left(U_{\lambda}(\beta,\theta)^{+}\right)^3\varphi_1-
\left(U_{\lambda}(\beta',\theta)^{+}\right)^3\varphi_1
\right\|\leq\\ &\leq&
4\times4\times4\times2(1+2+3+4+5)2\pi\left|\beta-\beta'\right|\\ &=&
2\times4^3(1+2+3+4+5)2\pi\left|\beta-\beta'\right|.
\end{eqnarray*} Inductively one finds that 
$\left(U_{\lambda}(\beta,\theta)^{+}\right)^n\varphi_1$ has at the most $4^n$ terms,
and according to
(\ref{StarEquation}) the largest contribution comes from the product
\[
e^{-i(1+2+\ldots+2n-3)2\pi\beta}e^{-i(((2n-2)+(2n-1))2\pi\beta)}=
e^{-i((1+2+\ldots+2n-1)2\pi\beta)}
\] and
then
\[\left\|\left(U_{\lambda}(\beta,\theta)^{+}\right)^n\varphi_1-
\left(U_{\lambda}(\beta',\theta)^{+}\right)^n\varphi_1\right\|
\]
\[
\leq2\times4^n(1+2+\ldots+2n-1)2\pi\left|\beta-\beta'\right|;\]  since
$2n^2-n=1+2+\ldots+2n-1$, the result follows.
\end{proof}

\

\subsection{Proof of Theorem~\ref{MainTheorem}(ii)} Finally, using this preparatory
set of results, we
finish the proof of our main result.

 Let $f(n)=(\ln(2+|n|))^{\frac{1}{5}}$. Sequences
$\beta_m,T_m,\Delta_m$ will be built inductively, starting with
$\beta_1=1$, so that
\begin{itemize}
\item[(i)] $\beta_{m+1}-\beta_m=2^{-\kappa_m!}$ for some
$\kappa_m\in\N$;

\item[(ii)]
$\frac{1}{T_m+1}\sum_{j=T_m}^{2T_m}\|P_{n\geq\frac{T_m}{f(T_m)}}
\left(U_{\lambda}(\beta,\theta)^{+}\right)^{j}\varphi_1\|^2\geq\frac{1}{f(T_m)^2}$
for all
$\theta\in[0,2\pi)$,
$\lambda\in[\frac{\pi}{6},\frac{\pi}{2}]$ and $\beta$ with
$|\beta-\beta_m|\leq\Delta_m$;

\item[(iii)] $|\beta_{m+1}-\beta_k|<\Delta_k$ for $k=1,2,\ldots,m$.
\end{itemize}
\

If (i), (ii) and (iii) are satisfied then we conclude by (i) that
$\beta_{\infty}=\lim\beta_m$ is irrational, by (iii) that
$|\beta_{\infty}-\beta_m|\leq\Delta_m$ and then by (ii) that
\[\frac{1}{T_m+1}\sum_{j=T_m}^{2T_m}\|P_{n\geq\frac{T_m}{f(T_m)}}
\left(U_{\lambda}(\beta_{\infty},\theta)^{+}\right)^{j}\varphi_1\|^2\geq\frac{1}{f(T_m)^2}\]
for
$\theta\in[0,2\pi)$ and
$\lambda\in[\frac{\pi}{6},\frac{\pi}{2}]$. So by Lemma
\ref{Lemma1}
\[\limsup_{n\rightarrow\infty}\|X\left(U_{\lambda}(\beta,\theta)^{+}\right)^{n}\varphi_1\|^2
\frac{f(n)^5}{n^2}=\infty\] for $\beta=\beta_{\infty}$ and the result is proved.

Then we shall construct $\beta_m,T_m,\Delta_m$ such that (i), (ii) and (iii) hold.
Start with
$\beta_1=1$. Given
$\beta_1,\ldots,\beta_m$, $T_1,\ldots,T_{m-1}$ and
$\Delta_1,\ldots,\Delta_{m-1}$ we shall show how to choose
$T_m,\Delta_m$ and $\beta_{m+1}$.

Given $\beta_m$, let $\varphi_1=\eta+\psi$ be the decomposition given by
Lemma~\ref{Lemma4} and let
$C_1,C_2$ be the corresponding constants. Choose $T_m\geq2T_{m-1}$ (and
$T_1\geq2$) so that
\begin{equation}\label{TheoremEquation}
C_1^2-3\sqrt{2\pi}C_2(2f(T_m)^{-1}+T_m^{-1})^{\frac{1}{2}}\geq2f(T_m)^{-1}.
\end{equation} This is possible since $C_1$ and $C_2$ are fixed (given
$\beta_m$) and
$f(n)\rightarrow\infty$.

Note that
\begin{equation}\label{TheoremEquation1}
\frac{1}{T+1}\sum_{j=T}^{2T}\|P_{n<
\frac{T}{f(T)}}\left(U_{\lambda}(\beta,\theta)^{+}\right)^j\psi\|^2\leq\frac{2\pi}{T+1}
\#\Big\{n:n<\frac{T}{f(T)}\Big\}\trip{\psi}^2;
\end{equation} in fact
\begin{eqnarray*} && \frac{1}{T+1}\sum_{j=T}^{2T}\|P_{n<
\frac{T}{f(T)}}\left(U_{\lambda}(\beta,\theta)^{+}\right)^j\psi\|^2=\\ &=&
\frac{1}{T+1}\sum_{j=T}^{2T}\sum_{n<
\frac{T}{f(T)}}\left|\left(\left(U_{\lambda}(\beta,\theta)^{+}\right)^j\psi\right)(n)
\right|^2\\ &=& \frac{1}{T+1}\sum_{n<
\frac{T}{f(T)}}\sum_{j=T}^{2T}\left|\left(\left(U_{\lambda}(\beta,\theta)^{+}\right)^j
\psi\right)(n)\right|^2\\ &\leq& \frac{1}{T+1}\sum_{n<
\frac{T}{f(T)}}\sum_{j=-\infty}^{\infty}
\left|\langle\varphi_n,\left(U_{\lambda}(\beta,\theta)^{+}\right)^j\psi\rangle\right|^2,
\end{eqnarray*} then by Lemma~\ref{Lemma3}
\[\frac{1}{T+1}\sum_{j=T}^{2T}\|P_{n<
\frac{T}{f(T)}}\left(U_{\lambda}(\beta,\theta)^{+}\right)^j\psi\|^2\leq\frac{1}{T+1}
\sum_{n< \frac{T}{f(T)}}2\pi\trip{\psi}^2,\] and (\ref{TheoremEquation1}) follows.

By Lemma~\ref{Lemma2} and (\ref{TheoremEquation1})
\begin{eqnarray*} &&\frac{1}{T_m+1}\sum_{j=T_m}^{2T_m}\|P_{n\geq
\frac{T_m}{f(T_m)}}\left(U_{\lambda}(\beta_m,\theta)^{+}\right)^j\varphi_1\|^2\geq\\
&\geq& \|\psi\|^2
-3\Big(\frac{1}{T_m+1}\sum_{j=T_m}^{2T_m}\|P_{n<
\frac{T_m}{f(T_m)}}\left(U_{\lambda}(\beta_m,\theta)^{+}\right)^j\psi\|^2
\Big)^{\frac{1}{2}}\\ &\geq&
\|\psi\|^2-3\Big(\frac{2\pi}{T_m+1}\#\Big\{n:n<\frac{T_m}{f(T_m)}\Big\}
\trip{\psi}^2\Big)^{\frac{1}{2}}\\ &\geq&
C_1^2-3\Big(\frac{2\pi}{T_m+1}\#\Big\{n:n<\frac{T_m}{f(T_m)}\Big\}
C_2^2\Big)^{\frac{1}{2}}\\ &=& C_1^2-3\sqrt{2\pi}C_2\Big(\frac{1}{T_m+1}
\#\Big\{n:n<\frac{T_m}{f(T_m)}\Big\}\Big)^{\frac{1}{2}}.
\end{eqnarray*} Since
$\#\Big\{n:n<\frac{T_m}{f(T_m)}\Big\}\leq2\frac{T_m}{f(T_m)}+1$ it follows that
\begin{eqnarray*} & &\frac{1}{T_m+1}\sum_{j=T_m}^{2T_m}\|P_{n\geq
\frac{T_m}{f(T_m)}}\left(U_{\lambda}(\beta_m,\theta)^{+}\right)^j\varphi_1\|^2\\ &\geq&
C_1^2-3\sqrt{2\pi}C_2\Big(\frac{1}{T_m+1}\Big(\frac{2T_m}{f(T_m)}
+1\Big)\Big)^{\frac{1}{2}}\\ &\geq&
C_1^2-3\sqrt{2\pi}C_2\Big(\frac{2}{f(T_m)}+\frac{1}{T_m}\Big)^{\frac{1}{2}}
\end{eqnarray*} for $\theta\in[0,2\pi)$ and
$\lambda\in[\frac{\pi}{6},\frac{\pi}{2}]$. Thus by (\ref{TheoremEquation}), we obtain
\[\frac{1}{T_m+1}\sum_{j=T_m}^{2T_m}\|P_{n\geq
\frac{T_m}{f(T_m)}}\left(U_{\lambda}(\beta_m,\theta)^{+}\right)^j\varphi_1\|^2
\geq\frac{2}{f(T_m)}\] for $\theta\in[0,2\pi)$ and
$\lambda\in[\frac{\pi}{6},\frac{\pi}{2}]$.

So, by Lemma~\ref{Lemma5}, for $\beta\in\R$, $\theta\in[0,2\pi)$ and
$\lambda\in[\frac{\pi}{6},\frac{\pi}{2}]$
\begin{eqnarray*} & & \frac{1}{T_m+1}\sum_{j=T_m}^{2T_m}\|P_{n\geq
\frac{T_m}{f(T_m)}}\left(U_{\lambda}(\beta,\theta)^{+}\right)^j\varphi_1\|^2\geq\\
&\geq&
\Big(\frac{1}{T_m+1}\sum_{j=T_m}^{2T_m}\|P_{|n|\geq
\frac{T_m}{f(T_m)}}\left(U_{\lambda}(\beta,\theta)^{+}\right)^j\varphi_1\|\Big)^2\\ &=&
\Bigg(\frac{1}{T_m+1}\sum_{j=T_m}^{2T_m}\Big\|P_{n\geq
\frac{T_m}{f(T_m)}}\left(U_{\lambda}(\beta_m,\theta)^{+}\right)^j\varphi_1\\ & & +
P_{n\geq
\frac{T_m}{f(T_m)}}\left(\left(U_{\lambda}(\beta,\theta)^{+}\right)^j\varphi_1
-\left(U_{\lambda}(\beta_m,\theta)^{+}\right)^j\varphi_1\right)\Big\|\Bigg)^2\\ &\geq&
\Bigg(\frac{1}{T_m+1}\sum_{j=T_m}^{2T_m}\left\|P_{n\geq
\frac{T_m}{f(T_m)}}\left(U_{\lambda}(\beta_m,\theta)^{+}\right)^j\varphi_1\right\|\\ &&
-\left\|\left(\left(U_{\lambda}(\beta,\theta)^{+}\right)^j-
\left(U_{\lambda}(\beta_m,\theta)^{+}\right)^j\right)\varphi_1\right\|\Bigg)^2\\ &\geq&
\Big(\frac{1}{T_m+1}\sum_{j=T_m}^{2T_m}(\|P_{n\geq
\frac{T_m}{f(T_m)}}\left(U_{\lambda}(\beta_m,\theta)^{+}\right)^j\varphi_1\|^2\\ &&
-4^{j+1}(2j^2-j)\pi
|\beta-\beta_m|)\Big)^2\\ &\geq&
\Big(\frac{2}{f(T_m)}-\frac{1}{T_m+1}\Big(\sum_{j=T_m}^{2T_m}4^{j+1}(2j^2-j)\pi\Big)
|\beta-\beta_m|\Big)^2.
\end{eqnarray*} Taking
\[\Delta_m=\frac{T_m+1}{f(T_m)\sum_{j=T_m}^{2T_m}4^{j+1}(2j^2-j)\pi}\] we obtain
that, if
$|\beta-\beta_m|<\Delta_m$,
\[\frac{1}{T_m+1}\sum_{j=T_m}^{2T_m}\|P_{n\geq
\frac{T_m}{f(T_m)}}\left(U_{\lambda}(\beta,\theta)^{+}\right)^j\varphi_1\|^2
\geq\frac{1}{f(T_m)^2}.\] Finally, pick $\beta_{m+1}$ rational so that
\[|\beta_n-\beta_{m+1}|<\Delta_n\qquad n=1,\ldots,m,\] and
$\beta_{m+1}=\beta_m+2^{-\kappa_m!}$ for some $\kappa_m\in\N$. This finishes the
proof of
Theorem~\ref{MainTheorem}(ii).

\

\noindent \textit{Remark.} For the operator
$U_{\lambda}(\beta,\theta):= U(\beta,\theta)(\Id+(e^{i\lambda}-1)P_{\varphi_1})$ on
$l^2(\Z)$ we can
similarly prove an analogous result. The proof of dynamical instability for some
irrational $\beta$ is
essentially unchanged except for Lemma~\ref{Lemma4} which is simplified since
$U(\beta,\theta)$ is purely absolutely continuous for $\beta$ rational. On the other
hand, about pure point
spectrum, the main difference in this case is that $\varphi_1$ might not be cyclic,
an thus, we don't get
pure point spectrum for
$U_{\lambda}(\beta,\theta)$ for a.e. $\theta$ and $\lambda$ as obtained on
$l^2(\N^*)$, but we get that
$\varphi_1$ is in the point spectral subspace corresponding to
$U_{\lambda}(\beta,\theta)$ for a.e. $\theta$ and $\lambda$.

\

\end{document}